\ifdefined\pdfoutput\pdfoutput=1\fi 

\documentclass[journal]{IEEEtran}

\usepackage[utf8]{inputenc}

\usepackage[T1]{fontenc}

\usepackage{textcomp}

\usepackage{amsmath}

\usepackage{graphicx}

\usepackage{booktabs}

\usepackage{array}

\usepackage{url}

\usepackage[hidelinks]{hyperref}

\usepackage{pgfplots}
\pgfplotsset{compat=1.16}
\usepgfplotslibrary{fillbetween}
\definecolor{ajnavy}{HTML}{1F3A5F}\definecolor{ajcoral}{HTML}{C8553D}\definecolor{ajgrey}{HTML}{8A8A8A}

\newcolumntype{L}[1]{>{\raggedright\arraybackslash}p{#1}}

\begin{document}

\title{Behavioral Differences Between Active and Inactive Users in Competitive Programming: A Rating-Restricted Analysis of Codeforces Activity}

\author{Azuad Islam Ruhan, Golam Mostofa Naeem, Rakibul Islam Rafi, Sherin Afrin Mim, Nazira Bani Opi, Dewan Fahad Chowdhury, and Md. Rejaul Korim Sadi\thanks{Azuad Islam Ruhan (corresponding author) is an undergraduate researcher in the Department of Computer Science and Engineering, Metropolitan University, Sylhet, Bangladesh. E-mail: ahnafruhan7@gmail.com\par Golam Mostofa Naeem is a Senior Lecturer in the Department of Computer Science and Engineering, Metropolitan University, Sylhet, Bangladesh. E-mail: gmnaeem7@gmail.com\par Rakibul Islam Rafi is an undergraduate student in the Department of Computer Science and Engineering, Metropolitan University, Sylhet, Bangladesh. E-mail: rakibulislamrafi36@gmail.com\par Sherin Afrin Mim is an undergraduate student in the Department of Computer Science and Engineering, Metropolitan University, Sylhet, Bangladesh. E-mail: sherinafrinmim6@gmail.com\par Nazira Bani Opi is an undergraduate student in the Department of Computer Science and Engineering, Metropolitan University, Sylhet, Bangladesh. E-mail: nazirabani2@gmail.com\par Dewan Fahad Chowdhury is an undergraduate student in the Department of Computer Science and Engineering, Metropolitan University, Sylhet, Bangladesh. E-mail: dewanfahadc@gmail.com\par Md. Rejaul Korim Sadi is an undergraduate researcher in the Department of Computer Science and Engineering, Metropolitan University, Sylhet, Bangladesh. E-mail: rejaulkorimsadi@gmail.com}}

\maketitle

\begin{abstract}We compared the recorded behavior of competitive programmers who became inactive on Codeforces with that of active users of similar skill. From 6,927 accounts collected through the public Codeforces application programming interface in June 2026, we removed 888 accounts whose retrieved submission histories were incomplete and restricted both groups to a peak rating below 1500. This left 3,421 active accounts and 908 accounts with no submission for at least three months, labeled attrition. The differences between the groups were small. Accounts labeled attrition accepted smaller shares of their attempts, overall and on mathematics and data-structure problems (rank-biserial r from $-$0.16 to $-$0.12), and solved fewer problems per rated contest (median 5.95 versus 7.80; r = $-$0.13); struggle time, upsolving and recovery after rating drops did not differ after correction for multiple testing. Two pre-specified models using eight features that do not depend on the length of the observation window reached cross-validated areas under the receiver operating characteristic curve of 0.596 (logistic regression) and 0.611 (random forest), with probabilities barely better than the base rate. Across random samples, stricter cleaning rules, other definitions of inactivity and matching on account tenure, the value ranged from 0.58 to 0.66. Using only activity from the 6 or 12 months ending three months before each user's last submission, it was 0.59 to 0.65, so a small difference precedes inactivity, but it is too weak to identify individual users at risk. The data and code are publicly archived, and one command regenerates every reported number.\end{abstract}


\section{Introduction}

Competitive programming (CP) is a widely used extra-curricular route for building algorithmic reasoning in computer science students [1], [2]. Platforms such as Codeforces supply immediate verdicts, timed contests and a public rating against which progress can be measured [3]. Problems arrive without hints, under fixed time and memory limits, and the learner decides alone what to try next [4].

Many learners who start do not stay. Dropout from introductory programming courses is well documented: students who leave most often report a lack of time or motivation and difficulty with the course [5], and learning to program is widely recognized as hard [6]. Educational data mining has modeled dropout from grade records and from the activity logs of massive open online courses [7], [8]. Voluntary competitive practice sits outside these settings: it happens off the transcript and has no deadline, whereas data-mining studies of programming have mostly examined courses [9].

This paper asks what the public activity records of a large platform can and cannot tell us about learners who stop. We address three questions. RQ1: Do users who become inactive on Codeforces differ in their recorded behavior from active users whose peak rating lies in the same range, and by how much? RQ2: How well do behavioral features that do not depend on the length of the observation window separate the two groups, and how sensitive is that answer to the choices made in the analysis: which active users are sampled, how incomplete records are cleaned, how inactivity is defined, and how long users have held their accounts? RQ3: Is a difference already present in the activity recorded before users became inactive?

The paper makes three contributions. First, it builds a rating-restricted cohort of 4,329 Codeforces accounts with a documented stage-by-stage flow and an audit of incomplete platform histories. Second, it reports the size of the behavioral differences, with confidence intervals, separating measures that are comparable between active and inactive users from measures that depend on how long each user was observed. Third, it shows how a modest classification signal behaves across analysis choices, including measures restricted to the months before inactivity, reporting discrimination, precision–recall performance and calibration rather than a single best score. The data and code are archived with a DOI and regenerate every number in the paper.

\section{Related Work}

\subsection{Educational Data Mining and Dropout}

Work on student retention has moved from predicting degree completion with demographic and first-year records [7], [10] to modeling dynamic behavior such as submission patterns and forum participation in online courses [8]. In introductory programming, data logged while students program predicted course performance better than traditional background factors [11], and data-mining techniques can identify students at risk of failure early in a course [12]. Most of this work studies enrolled students with a defined course end, where dropout is observed directly.

\subsection{Programming Platform Telemetry}

Studies of programming environments and online judges treat compilation errors, inter-submission gaps and wrong-answer rates as signals of difficulty [9], [11]. These logs record what learners did, not why. Voluntary competitive participation also differs from course work: there is no deadline, and disengagement unfolds over weeks or months, so the moment of quitting is not observed but has to be inferred from inactivity.

\subsection{Motivation and Persistence}

Motivational orientation and self-efficacy are established predictors of persistence [13], [14], and the adoption of programming contests has been described as a diffusion process driven by peers and perceived benefit [15]. None of these factors is visible in platform logs, which bounds what a log-based study can explain. Using learners' activity data also raises questions of agency and consent [16], which shaped our decision to publish only pseudonymized aggregates.

\section{Data and Methods}

\subsection{Data Collection}

Submission histories, contest records and profile data were retrieved from the public Codeforces application programming interface (API). The sampling frame was the platform's list of rated users (user.ratedList), from which a pool of 26,000 handles was drawn, stratified by current rating band in proportion to band size with at least 50 per band (seed 42). Accounts were retained only if they had at least five rated contests, a current rating between 1 and 2399 and at least one accepted submission; 6,927 accounts met these criteria. For each account the collection requested up to 10,000 submissions (user.status) and the full rating history (user.rating).

All measures refer to one observation time, $T_{\mathrm{obs}}$, the moment the feature file was built. The collection notebook did not record it; matching each account's rounded age in the released file to its registration time in the raw API records places it on 23 June 2026 between 13:41 and 13:44 UTC. Handles are replaced by a salted hash in every released file.

\subsection{Labels}

We call an account inactive when its last retrieved submission lies at least three months before $T_{\mathrm{obs}}$, where months are computed as elapsed days divided by 30 and the threshold is applied to the unrounded value (at least 90 days). Accounts that meet this condition and never reached a rating of 1500 are labeled attrition. The label records inactivity at one date; it does not establish that a user has stopped for good, and Section IV-D tests how the results change with longer inactivity thresholds. Accounts with a last submission less than 90 days before $T_{\mathrm{obs}}$ are labeled active.

The rating restriction applies to both groups. Inactive accounts that had reached 1500 (411) are excluded, because leaving after reaching an intermediate level may have other causes, and active accounts are likewise restricted to a peak rating below 1500. Without this restriction, a classifier could separate the groups partly by skill level, since the inactive group would be capped by construction and the active group would not. The threshold of 1500 is a study choice inside the Codeforces Specialist band (1400–1599); Section IV-D repeats the analysis with 1400 and 1600.

\subsection{Incomplete Histories}

Every rated contest requires at least one submission, so an account with fewer retrieved submissions than rated contests has an incomplete history. Such accounts, and accounts with more than 15 rated contests per month of submission span, were excluded. The rule removed 888 of the 6,927 accounts (12.8\%), all of them active at $T_{\mathrm{obs}}$. The removed accounts had a median of 7 retrieved submissions against a median of 24 rated contests. The collection requested complete histories in a single call, and its consistency check flagged only responses that were completely empty, so partial responses were not detected at collection time. The feature file does not show why the affected accounts are all active; the raw API responses are needed to establish the cause (Section V). Because the rule was defined after the data had been inspected, Section IV-D also reports three stricter rules.

\subsection{Behavioral Measures}

We use eight measures that do not depend on how long an account was observed. Six are taken directly from the API data: the shares of accepted attempts overall, on mathematics-tagged problems and on data-structure-tagged problems; mean struggle time, the time from the first attempt on a problem to its acceptance; upsolves per rated contest, accepted practice submissions on problems of contests the user took part in; and post-drop recovery, the mean rating change in the contest that follows a rating drop (0 for accounts without a drop). Two are constructed:

\begin{equation}\mathrm{SPC}_{i} = S_{i} / C_{i}\label{eq:1}\end{equation}

\noindent where $S_{i}$ is the number of distinct problems user $i$ solved and $C_{i}$ the number of rated contests entered, and

\begin{equation}\mathrm{EDI}_{i} = U_{i} \times \Delta T_{\mathrm{struggle},i}\label{eq:2}\end{equation}

\noindent where $U_{i}$ is upsolves per rated contest and $\Delta T_{\mathrm{struggle},i}$ is mean struggle time in minutes. Problems solved per contest (SPC) relates practice volume to contest participation. The Editorial Dependency Index (EDI) was constructed to capture slow, after-the-fact solving. Four further measures, rated contests per month, problems solved per month, account age and the total number of rated contests, depend on the length of the observed activity span, which ends at the last submission for inactive users and at $T_{\mathrm{obs}}$ for active users. They are reported descriptively but are not used in the models. Table V in Appendix A gives every definition as implemented.

\subsection{Statistical Analysis}

Group differences are reported as medians with interquartile ranges and as rank-biserial correlations [17], r = P($X_{\mathrm{attrition}}$ > $X_{\mathrm{active}}$) $-$ P($X_{\mathrm{attrition}}$ < $X_{\mathrm{active}}$), so that a negative r means lower values in the attrition group. Confidence intervals come from 1,000 bootstrap resamples of accounts [18]. Table II reports both the Mann–Whitney p-value [19] and the Holm-adjusted p-value over the 13 measures tested [20], labeled as such.

Two models were specified before the analysis: logistic regression and a random forest with 200 trees [21], both implemented in scikit-learn [22] with median imputation and standardization fitted inside each training fold. Using all eligible accounts avoids depending on one random sample of active users. Performance is estimated by stratified 5-fold cross-validation repeated five times [23]. The main measure is the area under the receiver operating characteristic curve (ROC-AUC), with a 95\% confidence interval from 2,000 bootstrap resamples of the out-of-fold predictions; every metric is computed within each repeat and then averaged over the five repeats. Because only 21\% of accounts are labeled attrition, a class imbalance under which ROC-AUC alone can give a favorable impression [24], we also report the area under the precision–recall curve (PR-AUC) [25], whose value for random guessing equals that share, and the Brier score with a calibration curve [26], [27]. Restricting the analysis to two models limits the optimism that comes from selecting the best of many [28]. Features that leak information about the label from outside the observation period were excluded by design [29].

To test how much the results depend on analysis choices, we repeat the models (i) on 50 balanced samples of active users, (ii) under stricter rules for incomplete histories, (iii) with inactivity thresholds of 3, 6 and 12 months crossed with rating limits of 1400, 1500 and 1600, and (iv) after matching each attrition account to the active account nearest in account age and number of rated contests. The contribution of the two constructed measures is tested by removing them and comparing paired cross-validation folds. Permutation importance is computed on every held-out fold; it is reported with caution because several measures are correlated [30].

\section{Results}

To test whether the difference is present before users become inactive (RQ3), the eight model measures were recomputed from a window that ends three months before each account's last submission and covers the preceding 6 or 12 months; active accounts, whose last submission lies close to $T_{\mathrm{obs}}$, are treated in the same way. Accounts need at least five rated contests and one accepted submission inside the window. For this analysis the submission and rating histories of the cohort were retrieved again from the public API in November 2026, at a rate within the platform's limits, and cut at $T_{\mathrm{obs}}$. Histories were obtained for all 908 attrition accounts and for 2,669 of the 3,421 active accounts (78\%) before the retrieval was stopped; on these accounts the whole-history models give ROC-AUC 0.597 and 0.608, as on the full cohort (0.596 and 0.611), so the subset does not change the result. For the 908 attrition accounts, which made no submissions after their retrieval in June, the measures recomputed from these histories equal the released values for at least 99.9\% of accounts on every measure, so the retrieval reproduces the data analyzed above. The same two models and cross-validation are used.

\subsection{Cohort}

Table I gives the count at every stage. The analysis cohort contains 4,329 accounts: 3,421 active and 908 labeled attrition. The rating restriction leaves the two groups with nearly identical peak ratings (median 1153 versus 1154; standardized mean difference 0.05). The groups differ in tenure: the active accounts are younger (median account age 18.7 versus 33.5 months) but have entered the same median number of rated contests (13).

\begin{table}[!t]\centering\caption{Cohort construction, stage by stage}\label{tab:1}

\footnotesize\setlength{\tabcolsep}{4pt}

\begin{tabular}{@{}L{0.795\dimexpr \columnwidth-2\tabcolsep\relax}L{0.205\dimexpr \columnwidth-2\tabcolsep\relax}@{}}

\toprule Stage & Accounts \\ \midrule

Retrieved from the API ($\geq$ 5 rated contests, current rating 1–2399) & 6,927 \\

Peak rating below 2400 & 6,892 \\

Complete submission history (data-quality rule) & 6,012 \\

\quad active at $T_{\mathrm{obs}}$ & 4,693 \\

\quad inactive for $\geq$ 3 months at $T_{\mathrm{obs}}$ & 1,319 \\

Peak rating below 1500: active & 3,421 \\

Peak rating below 1500: inactive (labeled attrition) & 908 \\

Analysis cohort & 4,329 \\

\bottomrule\end{tabular}

\par\vspace{3pt}\begin{minipage}{\columnwidth}\raggedright\scriptsize The data-quality rule removes 888 accounts from the full file and 880 after the peak-rating step. Inactive accounts with a peak rating of 1500 or more (411) and active accounts with a peak of 1500 or more (1,272) are excluded.\end{minipage}

\end{table}

\subsection{Group Differences}

Table II and Fig. 1 report all measures. Among the eight measures used in the models, the largest differences are in acceptance: accounts labeled attrition accepted smaller shares of their attempts on mathematics-tagged problems (median 0.464 versus 0.500; r = $-$0.16, 95\% CI [$-$0.20, $-$0.12]), overall (r = $-$0.13) and on data-structure-tagged problems (r = $-$0.12). They also solved fewer problems per rated contest, a median of 5.95 against 7.80 (r = $-$0.13, 95\% CI [$-$0.17, $-$0.09]). Struggle time, upsolving, post-drop recovery and the Editorial Dependency Index do not differ after Holm correction. Every effect is small: no measure used in the models exceeds $|r|$ = 0.16.

The window-dependent measures show larger differences, partly by construction. Accounts labeled attrition entered fewer rated contests per month (median 0.89 versus 1.13) and solved fewer problems per month (5.35 versus 8.69). Their accounts are older, so their activity is spread over a longer span, while the total number of rated contests is similar. These measures are therefore descriptive only.

\begin{table*}[!t]\centering\caption{Behavioral measures, active (n = 3,421) versus attrition (n = 908)}\label{tab:2}

\footnotesize\setlength{\tabcolsep}{4pt}

\begin{tabular}{@{}L{0.284\dimexpr \textwidth-10\tabcolsep\relax}L{0.184\dimexpr \textwidth-10\tabcolsep\relax}L{0.184\dimexpr \textwidth-10\tabcolsep\relax}L{0.158\dimexpr \textwidth-10\tabcolsep\relax}L{0.095\dimexpr \textwidth-10\tabcolsep\relax}L{0.095\dimexpr \textwidth-10\tabcolsep\relax}@{}}

\toprule Measure & Active, median [IQR] & Attrition, median [IQR] & r [95\% CI] & p (unadj.) & p (Holm) \\ \midrule

Problems solved per contest & 7.8 [4.0, 13.3] & 6.0 [3.2, 10.9] & $-$0.13 [$-$0.17, $-$0.09] & < 0.001 & < 0.001 \\

Math-tag acceptance rate & 0.500 [0.417, 0.591] & 0.464 [0.380, 0.550] & $-$0.16 [$-$0.20, $-$0.12] & < 0.001 & < 0.001 \\

Data-structure acceptance rate & 0.436 [0.308, 0.571] & 0.397 [0.257, 0.500] & $-$0.12 [$-$0.16, $-$0.08] & < 0.001 & < 0.001 \\

Overall acceptance rate & 0.490 [0.413, 0.576] & 0.462 [0.382, 0.549] & $-$0.13 [$-$0.17, $-$0.09] & < 0.001 & < 0.001 \\

Struggle time (min) & 469 [36, 2779] & 424 [23, 3010] & $-$0.01 [$-$0.05, 0.03] & 0.539 & 1.000 \\

Upsolves per contest & 0.250 [0.000, 0.600] & 0.253 [0.000, 0.630] & 0.01 [$-$0.03, 0.06] & 0.540 & 1.000 \\

Post-drop recovery & 1.3 [0.0, 19.0] & 0.0 [$-$2.4, 16.8] & $-$0.05 [$-$0.09, $-$0.01] & 0.011 & 0.055 \\

Editorial dependency index & 82.6 [0.0, 1126.0] & 73.9 [0.0, 1177.5] & $-$0.00 [$-$0.04, 0.04] & 0.922 & 1.000 \\

Rated contests per month\textsuperscript{\dag} & 1.1 [0.6, 2.0] & 0.9 [0.5, 1.6] & $-$0.17 [$-$0.21, $-$0.13] & < 0.001 & < 0.001 \\

Problems solved per month\textsuperscript{\dag} & 8.7 [4.0, 17.2] & 5.3 [2.6, 10.7] & $-$0.26 [$-$0.30, $-$0.22] & < 0.001 & < 0.001 \\

Account age (months)\textsuperscript{\dag} & 18.7 [10.1, 29.7] & 33.5 [22.1, 51.0] & 0.48 [0.44, 0.51] & < 0.001 & < 0.001 \\

Rated contests (total)\textsuperscript{\dag} & 13.0 [8.0, 24.0] & 13.0 [7.0, 25.0] & 0.01 [$-$0.03, 0.05] & 0.627 & 1.000 \\

Submissions per contest\textsuperscript{\ddag} & 17.2 [9.0, 29.1] & 14.2 [8.0, 24.7] & $-$0.10 [$-$0.14, $-$0.05] & < 0.001 & < 0.001 \\

\bottomrule\end{tabular}

\par\vspace{3pt}\begin{minipage}{\textwidth}\raggedright\scriptsize r: rank-biserial correlation, P(attrition > active) $-$ P(attrition < active); negative values mean lower values in the attrition group. p (unadj.): two-sided Mann–Whitney test; p (Holm): adjusted over the 13 measures. \textsuperscript{\dag} Depends on the length of the observed activity span; descriptive only, not used in the models. \textsuperscript{\ddag} Data-quality diagnostic (Section IV-D).\end{minipage}

\end{table*}

\begin{figure}[!t]\centering
\begin{tikzpicture}
\begin{axis}[width=0.62\columnwidth, scale only axis, axis lines*=left, tick label style={font=\scriptsize}, label style={font=\scriptsize}, xlabel style={align=center}, legend style={font=\scriptsize, draw=none, legend columns=2, /tikz/every even column/.append style={column sep=6pt}}, legend cell align=left, clip=false, scaled x ticks=false, scaled y ticks=false, height=6.2cm, xmin=-0.3, xmax=0.55, ymin=-0.7, ymax=11.7, ytick={0,...,11}, yticklabels={{Problems solved per month},{Rated contests per month},{Math-tag acceptance rate},{Overall acceptance rate},{Problems solved per contest},{Data-structure acceptance rate},{Post-drop recovery},{Struggle time (min)},{Editorial dependency index},{Rated contests (total)},{Upsolves per contest},{Account age (months)}},
 xlabel={Rank-biserial $r$ with 95\% bootstrap CI\\(negative: lower in the attrition group)}, legend style={at={(0.25,-0.22)}, anchor=north, legend columns=1}]
\draw[black, line width=0.5pt] (axis cs:0,-0.7) -- (axis cs:0,11.7);
\addplot+[only marks, mark=*, mark size=1.8pt, color=ajnavy, error bars/.cd, x dir=both, x explicit, error bar style={line width=0.7pt}, error mark=none] coordinates {(-0.1607,2.0000) += (0.0444,0) -= (0.0412,0) (-0.1333,3.0000) += (0.0436,0) -= (0.0382,0) (-0.1309,4.0000) += (0.0400,0) -= (0.0425,0) (-0.1229,5.0000) += (0.0417,0) -= (0.0409,0) (-0.0544,6.0000) += (0.0426,0) -= (0.0396,0) (-0.0132,7.0000) += (0.0426,0) -= (0.0413,0) (-0.0021,8.0000) += (0.0425,0) -= (0.0410,0) (0.0131,10.0000) += (0.0436,0) -= (0.0419,0)};
\addlegendentry{Used in the models}
\addplot+[only marks, mark=o, mark size=1.8pt, color=ajgrey, mark options={fill=white}, error bars/.cd, x dir=both, x explicit, error bar style={line width=0.7pt}, error mark=none] coordinates {(-0.2578,0.0000) += (0.0405,0) -= (0.0378,0) (-0.1695,1.0000) += (0.0394,0) -= (0.0393,0) (0.0105,9.0000) += (0.0434,0) -= (0.0415,0) (0.4754,11.0000) += (0.0337,0) -= (0.0377,0)};
\addlegendentry{Descriptive only (depends on window)}
\end{axis}
\end{tikzpicture}
\caption{Rank-biserial correlations with 95\% bootstrap confidence intervals for the twelve behavioral measures. Filled markers: measures used in the models; open markers: measures that depend on the length of the observed activity span.}\label{fig:fig1}\end{figure}

\subsection{Model Performance}

Table III summarizes performance. Logistic regression reached a cross-validated ROC-AUC of 0.596 (95\% CI [0.574, 0.617]) and the random forest 0.611 ([0.590, 0.631]). Both are above chance, but the separation is weak. The PR-AUC values (0.286 and 0.295) exceed the base rate of 0.210, and the Brier scores (0.163 and 0.164) are close to the value of 0.166 obtained by predicting the base rate for every account. In Fig. 2 the mean predicted probability per decile ranges from 0.07 to 0.45 for the random forest and from 0.11 to 0.32 for logistic regression; the random forest underestimates the share of attrition in its lowest deciles and overestimates it in its highest decile. The models rank accounts slightly better than chance but do not provide usable individual risk estimates.

\begin{table*}[!t]\centering\caption{Cross-validated performance of the two pre-specified models (n = 4,329)}\label{tab:3}

\footnotesize\setlength{\tabcolsep}{4pt}

\begin{tabular}{@{}L{0.232\dimexpr \textwidth-10\tabcolsep\relax}L{0.202\dimexpr \textwidth-10\tabcolsep\relax}L{0.091\dimexpr \textwidth-10\tabcolsep\relax}L{0.202\dimexpr \textwidth-10\tabcolsep\relax}L{0.121\dimexpr \textwidth-10\tabcolsep\relax}L{0.152\dimexpr \textwidth-10\tabcolsep\relax}@{}}

\toprule Model & ROC-AUC [95\% CI] & Fold SD & PR-AUC [95\% CI] & Brier score & Balanced accuracy \\ \midrule

Logistic regression & 0.596 [0.574, 0.617] & 0.024 & 0.286 [0.263, 0.313] & 0.163 & 0.562 \\

Random forest & 0.611 [0.590, 0.631] & 0.019 & 0.295 [0.269, 0.322] & 0.164 & 0.564 \\

Reference: chance / base rate & 0.500 & – & 0.210 & 0.166 & 0.500 \\

\bottomrule\end{tabular}

\par\vspace{3pt}\begin{minipage}{\textwidth}\raggedright\scriptsize Stratified 5-fold cross-validation repeated five times on all eligible accounts; each metric is computed on the out-of-fold predictions of one repeat and averaged over the five repeats. Confidence intervals: 2,000 bootstrap resamples of accounts in each repeat, bounds averaged over repeats. Balanced accuracy at a threshold that flags the same share of accounts as the base rate. The base-rate Brier score is p(1 $-$ p) with p the share of attrition.\end{minipage}

\end{table*}

\begin{figure}[!t]\centering
\begin{tikzpicture}
\begin{axis}[width=0.78\columnwidth, scale only axis, axis lines*=left, tick label style={font=\scriptsize}, label style={font=\scriptsize}, xlabel style={align=center}, legend style={font=\scriptsize, draw=none, legend columns=2, /tikz/every even column/.append style={column sep=6pt}}, legend cell align=left, clip=false, scaled x ticks=false, scaled y ticks=false, height=4.6cm, xmin=0, xmax=0.6, ymin=0, ymax=0.6,
 xlabel={Predicted probability of attrition (decile mean)}, ylabel={Observed share of attrition},
 legend style={at={(0.02,0.98)}, anchor=north west, legend columns=1}]
\addplot[black, dashed, forget plot] coordinates {(0,0) (0.6,0.6)};
\addplot[ajgrey, dotted, forget plot] coordinates {(0,0.2097) (0.6,0.2097)};
\node[anchor=south east, font=\scriptsize, text=ajgrey] at (axis cs:0.6,0.2097) {base rate 0.21};
\addplot[name path=lrlo, draw=none, forget plot] coordinates {(0.1106,0.1211) (0.1510,0.1099) (0.1705,0.1368) (0.1867,0.1302) (0.2016,0.1533) (0.2148,0.1719) (0.2293,0.1871) (0.2461,0.1784) (0.2675,0.2173) (0.3178,0.3183)};
\addplot[name path=lrhi, draw=none, forget plot] coordinates {(0.1106,0.1899) (0.1510,0.1744) (0.1705,0.2095) (0.1867,0.2010) (0.2016,0.2283) (0.2148,0.2500) (0.2293,0.2613) (0.2461,0.2565) (0.2675,0.3009) (0.3178,0.4072)};
\addplot[ajnavy, fill opacity=0.18, draw=none, forget plot] fill between[of=lrhi and lrlo];
\addplot[color=ajnavy, mark=*, mark size=1.2pt, line width=0.8pt] coordinates {(0.1106,0.1547) (0.1510,0.1409) (0.1705,0.1732) (0.1867,0.1640) (0.2016,0.1898) (0.2148,0.2102) (0.2293,0.2263) (0.2461,0.2148) (0.2675,0.2587) (0.3178,0.3649)};
\addlegendentry{Logistic regression}
\addplot[name path=rflo, draw=none, forget plot] coordinates {(0.0619,0.0907) (0.1028,0.1000) (0.1329,0.1206) (0.1595,0.1232) (0.1872,0.1535) (0.2165,0.1626) (0.2498,0.2022) (0.2868,0.2209) (0.3375,0.2370) (0.4449,0.3103)};
\addplot[name path=rfhi, draw=none, forget plot] coordinates {(0.0619,0.1524) (0.1028,0.1654) (0.1329,0.1868) (0.1595,0.1903) (0.1872,0.2259) (0.2165,0.2449) (0.2498,0.2800) (0.2868,0.3002) (0.3375,0.3227) (0.4449,0.3949)};
\addplot[ajcoral, fill opacity=0.18, draw=none, forget plot] fill between[of=rfhi and rflo];
\addplot[color=ajcoral, mark=*, mark size=1.2pt, line width=0.8pt] coordinates {(0.0619,0.1200) (0.1028,0.1313) (0.1329,0.1511) (0.1595,0.1548) (0.1872,0.1879) (0.2165,0.2050) (0.2498,0.2399) (0.2868,0.2597) (0.3375,0.2799) (0.4449,0.3527)};
\addlegendentry{Random forest}
\end{axis}
\end{tikzpicture}
\caption{Calibration of the out-of-fold predicted probabilities of the first cross-validation repeat (deciles) with 95\% bootstrap bands. The dotted line marks the base rate of attrition.}\label{fig:fig2}\end{figure}

\subsection{Robustness to Analysis Choices}

Table IV and Fig. 3 repeat the models under different choices. On 50 balanced samples of 908 active accounts, the random forest averaged 0.598 (2.5th–97.5th percentile 0.569–0.627) and logistic regression 0.593 (0.578–0.610). A result reported from a single sample can therefore differ from the typical value by several hundredths.

Among the accounts kept by the data-quality rule, 117 active accounts but only 4 attrition accounts have fewer than two submissions per rated contest; the medians are 17.20 and 14.15. Some incomplete histories may therefore remain among active accounts. Requiring at least 2, 3 or 5 submissions per contest changes the random-forest ROC-AUC to 0.597, 0.599 and 0.606, and the logistic-regression value to 0.597, 0.597 and 0.597.

Longer inactivity thresholds give similar or slightly higher values: at 12 months with the 1500 limit, 0.624 for the random forest and 0.656 for logistic regression, with 282 accounts labeled attrition. Rating limits of 1400 and 1600 leave the results essentially unchanged. The difference in problems solved per contest keeps its direction in all nine definitions (r from $-$0.19 to $-$0.12).

Matching each attrition account to the active account nearest in account age and number of rated contests removes the tenure difference (median age of the matched active accounts 33.2 months). The ROC-AUC then falls to 0.579 for the random forest (range over 20 matchings 0.566–0.598) and 0.591 for logistic regression. Part of the signal is therefore associated with tenure. Across all the analyses in Table IV the values stay between 0.58 and 0.66: a small but persistent difference, not a basis for identifying individual users.

\begin{table*}[!t]\centering\caption{ROC-AUC under alternative analysis choices}\label{tab:4}

\footnotesize\setlength{\tabcolsep}{4pt}

\begin{tabular}{@{}L{0.424\dimexpr \textwidth-6\tabcolsep\relax}L{0.111\dimexpr \textwidth-6\tabcolsep\relax}L{0.232\dimexpr \textwidth-6\tabcolsep\relax}L{0.232\dimexpr \textwidth-6\tabcolsep\relax}@{}}

\toprule Analysis & Accounts & Logistic regression & Random forest \\ \midrule

Primary: all eligible accounts & 4,329 & 0.596 [0.574, 0.617] & 0.611 [0.590, 0.631] \\

Balanced samples (50 draws), mean [2.5–97.5\%] & 1,816 & 0.593 [0.578, 0.610] & 0.598 [0.569, 0.627] \\

$\geq$ 2 submissions per contest & 4,208 & 0.597 $\pm$ 0.012 & 0.597 $\pm$ 0.022 \\

$\geq$ 3 submissions per contest & 4,118 & 0.597 $\pm$ 0.022 & 0.599 $\pm$ 0.016 \\

$\geq$ 5 submissions per contest & 3,882 & 0.597 $\pm$ 0.030 & 0.606 $\pm$ 0.022 \\

Peak < 1400, inactive $\geq$ 3 months & 3,724 & 0.592 $\pm$ 0.018 & 0.609 $\pm$ 0.017 \\

Peak < 1400, inactive $\geq$ 6 months & 3,402 & 0.628 $\pm$ 0.034 & 0.621 $\pm$ 0.021 \\

Peak < 1400, inactive $\geq$ 12 months & 3,178 & 0.656 $\pm$ 0.021 & 0.620 $\pm$ 0.022 \\

Peak < 1500, inactive $\geq$ 6 months & 3,966 & 0.634 $\pm$ 0.040 & 0.630 $\pm$ 0.025 \\

Peak < 1500, inactive $\geq$ 12 months & 3,703 & 0.656 $\pm$ 0.019 & 0.624 $\pm$ 0.023 \\

Peak < 1600, inactive $\geq$ 3 months & 4,701 & 0.597 $\pm$ 0.013 & 0.610 $\pm$ 0.016 \\

Peak < 1600, inactive $\geq$ 6 months & 4,314 & 0.635 $\pm$ 0.012 & 0.624 $\pm$ 0.017 \\

Peak < 1600, inactive $\geq$ 12 months & 4,027 & 0.660 $\pm$ 0.029 & 0.615 $\pm$ 0.040 \\

Matched on account age and contests (20 matchings), mean [range] & 1,816 & 0.591 [0.583, 0.599] & 0.579 [0.566, 0.598] \\

Before inactivity: 6-month window ending 3 months before the last submission & 1,506 & 0.650 [0.615, 0.684] & 0.624 [0.587, 0.661] \\

Before inactivity: 12-month window ending 3 months before the last submission & 2,091 & 0.609 [0.580, 0.638] & 0.586 [0.557, 0.616] \\

\bottomrule\end{tabular}

\par\vspace{3pt}\begin{minipage}{\textwidth}\raggedright\scriptsize $\pm$ values are standard deviations over cross-validation folds; brackets give 95\% bootstrap confidence intervals unless stated. Rows with a rating limit or inactivity threshold change only that element of the primary definition (peak < 1500, inactive $\geq$ 3 months). The two last rows use only accounts with at least five rated contests in the window.\end{minipage}

\end{table*}

\begin{figure}[!t]\centering
\begin{tikzpicture}
\begin{axis}[width=0.5\columnwidth, scale only axis, axis lines*=left, tick label style={font=\scriptsize}, label style={font=\scriptsize}, xlabel style={align=center}, legend style={font=\scriptsize, draw=none, legend columns=2, /tikz/every even column/.append style={column sep=6pt}}, legend cell align=left, clip=false, scaled x ticks=false, scaled y ticks=false, height=7.2cm, xmin=0.48, xmax=0.7, ymin=-0.7, ymax=13.7, ytick={0,...,13},
 yticklabels={{Peak < 1600, inactive $\geq$ 12 months},{Peak < 1600, inactive $\geq$ 6 months},{Peak < 1600, inactive $\geq$ 3 months},{Peak < 1500, inactive $\geq$ 12 months},{Peak < 1500, inactive $\geq$ 6 months},{Peak < 1400, inactive $\geq$ 12 months},{Peak < 1400, inactive $\geq$ 6 months},{Peak < 1400, inactive $\geq$ 3 months},{Matched on account age and contests},{At least 5 submissions per contest},{At least 3 submissions per contest},{At least 2 submissions per contest},{Balanced samples (50 draws)},{Primary: all eligible accounts}}, xlabel={ROC-AUC (0.5 = chance)}, xtick={0.5,0.55,0.6,0.65,0.7},
 x tick label style={/pgf/number format/fixed, /pgf/number format/precision=2}, legend style={at={(0.2,-0.13)}, anchor=north}]
\draw[black, dashed, line width=0.5pt] (axis cs:0.5,-0.7) -- (axis cs:0.5,13.7);
\addplot+[only marks, mark=*, mark size=1.5pt, color=ajnavy, error bars/.cd, x dir=both, x explicit, error bar style={line width=0.7pt}, error mark=none] coordinates {(0.5959,13.1600) += (0.0208,0) -= (0.0216,0) (0.5926,12.1600) += (0.0173,0) -= (0.0142,0) (0.5966,11.1600) += (0.0123,0) -= (0.0123,0) (0.5975,10.1600) += (0.0216,0) -= (0.0216,0) (0.5971,9.1600) += (0.0296,0) -= (0.0296,0) (0.5908,8.1600) += (0.0084,0) -= (0.0082,0) (0.5918,7.1600) += (0.0177,0) -= (0.0177,0) (0.6283,6.1600) += (0.0338,0) -= (0.0338,0) (0.6557,5.1600) += (0.0211,0) -= (0.0211,0) (0.6337,4.1600) += (0.0403,0) -= (0.0403,0) (0.6564,3.1600) += (0.0186,0) -= (0.0186,0) (0.5974,2.1600) += (0.0134,0) -= (0.0134,0) (0.6350,1.1600) += (0.0124,0) -= (0.0124,0) (0.6605,0.1600) += (0.0290,0) -= (0.0290,0)};
\addlegendentry{Logistic regression}
\addplot+[only marks, mark=square*, mark size=1.4pt, color=ajcoral, error bars/.cd, x dir=both, x explicit, error bar style={line width=0.7pt}, error mark=none] coordinates {(0.6113,12.8400) += (0.0196,0) -= (0.0211,0) (0.5975,11.8400) += (0.0296,0) -= (0.0288,0) (0.5970,10.8400) += (0.0225,0) -= (0.0225,0) (0.5991,9.8400) += (0.0161,0) -= (0.0161,0) (0.6060,8.8400) += (0.0216,0) -= (0.0216,0) (0.5786,7.8400) += (0.0196,0) -= (0.0131,0) (0.6093,6.8400) += (0.0167,0) -= (0.0167,0) (0.6208,5.8400) += (0.0213,0) -= (0.0213,0) (0.6196,4.8400) += (0.0223,0) -= (0.0223,0) (0.6297,3.8400) += (0.0254,0) -= (0.0254,0) (0.6237,2.8400) += (0.0231,0) -= (0.0231,0) (0.6101,1.8400) += (0.0157,0) -= (0.0157,0) (0.6237,0.8400) += (0.0174,0) -= (0.0174,0) (0.6153,-0.1600) += (0.0396,0) -= (0.0396,0)};
\addlegendentry{Random forest}
\end{axis}
\end{tikzpicture}
\caption{ROC-AUC of logistic regression (circles) and random forest (squares) under the analyses of Table IV, with 95\% confidence intervals (primary), 2.5–97.5\% ranges (balanced samples), ranges over matchings, or $\pm$ one fold standard deviation. The dashed line marks chance.}\label{fig:fig3}\end{figure}

\subsection{Contribution of Individual Measures}

Removing the Editorial Dependency Index changes the ROC-AUC by +0.001 for logistic regression and $-$0.004 for the random forest, so it adds nothing measurable. Removing problems solved per contest lowers the values by 0.015 and 0.033 (Fig. 4(b)). Problems solved per contest also has the largest permutation importance in the random forest (mean drop 0.050; 2.5th–97.5th percentile over folds 0.023–0.077) and ranks first in 48\% of folds (Fig. 4(a)). Several measures are strongly correlated (for example, the overall and mathematics-tagged acceptance rates), and permutation importance can misattribute importance among correlated measures [30], so the ranking below the first measure is not reliable.

\begin{figure}[!t]\centering
\begin{tikzpicture}
\begin{axis}[width=0.52\columnwidth, scale only axis, axis lines*=left, tick label style={font=\scriptsize}, label style={font=\scriptsize}, xlabel style={align=center}, legend style={font=\scriptsize, draw=none, legend columns=2, /tikz/every even column/.append style={column sep=6pt}}, legend cell align=left, clip=false, scaled x ticks=false, scaled y ticks=false, height=4.4cm, xmin=-0.0156, xmax=0.0824, ymin=-0.7, ymax=7.7, ytick={0,...,7},
 yticklabels={{Post-drop recovery},{Editorial dependency index},{Data-structure acceptance rate},{Overall acceptance rate},{Upsolves per contest},{Struggle time (min)},{Math-tag acceptance rate},{Problems solved per contest}}, xlabel={Drop in ROC-AUC when permuted (random forest;\\mean and 2.5--97.5\% over folds)},
 x tick label style={/pgf/number format/fixed, /pgf/number format/precision=2}]
\draw[black, line width=0.5pt] (axis cs:0,-0.7) -- (axis cs:0,7.7);
\addplot+[only marks, mark=*, mark size=1.5pt, color=ajnavy, error bars/.cd, x dir=both, x explicit, error bar style={line width=0.7pt}, error mark=none] coordinates {(0.0060,0.0000) += (0.0157,0) -= (0.0166,0) (0.0100,1.0000) += (0.0185,0) -= (0.0164,0) (0.0155,2.0000) += (0.0181,0) -= (0.0169,0) (0.0157,3.0000) += (0.0218,0) -= (0.0229,0) (0.0214,4.0000) += (0.0189,0) -= (0.0174,0) (0.0322,5.0000) += (0.0283,0) -= (0.0265,0) (0.0459,6.0000) += (0.0293,0) -= (0.0307,0) (0.0501,7.0000) += (0.0272,0) -= (0.0272,0)};
\end{axis}
\end{tikzpicture}\\[-1pt]{\footnotesize (a)}\\[4pt]
\begin{tikzpicture}
\begin{axis}[width=0.52\columnwidth, scale only axis, axis lines*=left, tick label style={font=\scriptsize}, label style={font=\scriptsize}, xlabel style={align=center}, legend style={font=\scriptsize, draw=none, legend columns=2, /tikz/every even column/.append style={column sep=6pt}}, legend cell align=left, clip=false, scaled x ticks=false, scaled y ticks=false, height=2.4cm, xmin=-0.0701, xmax=0.0156, ymin=-0.6, ymax=2.6, ytick={0,...,2},
 yticklabels={{Without both},{Without problems per contest},{Without EDI}}, xlabel={Change in ROC-AUC vs.\ all 8 features\\(mean and 2.5--97.5\% over paired folds)},
 x tick label style={/pgf/number format/fixed, /pgf/number format/precision=2}, legend style={at={(0.2,-0.62)}, anchor=north}]
\draw[black, line width=0.5pt] (axis cs:0,-0.6) -- (axis cs:0,2.6);
\addplot+[only marks, mark=*, mark size=1.5pt, color=ajnavy, error bars/.cd, x dir=both, x explicit, error bar style={line width=0.7pt}, error mark=none] coordinates {(0.0009,2.1400) += (0.0050,0) -= (0.0021,0) (-0.0146,1.1400) += (0.0167,0) -= (0.0137,0) (-0.0137,0.1400) += (0.0200,0) -= (0.0146,0)};
\addlegendentry{Logistic regression}
\addplot+[only marks, mark=square*, mark size=1.4pt, color=ajcoral, error bars/.cd, x dir=both, x explicit, error bar style={line width=0.7pt}, error mark=none] coordinates {(-0.0042,1.8600) += (0.0148,0) -= (0.0256,0) (-0.0334,0.8600) += (0.0215,0) -= (0.0305,0) (-0.0345,-0.1400) += (0.0223,0) -= (0.0306,0)};
\addlegendentry{Random forest}
\end{axis}
\end{tikzpicture}\\[-1pt]{\footnotesize (b)}
\caption{(a) Permutation importance in the random forest on every held-out fold (mean and 2.5–97.5\% range). (b) Change in ROC-AUC when the constructed measures are removed, over paired cross-validation folds.}\label{fig:fig4}\end{figure}

\subsection{Activity Before Inactivity}

In the window of 6 months, 1,208 active and 298 attrition accounts had at least five rated contests and one accepted submission; the ROC-AUC was 0.650 [0.615, 0.684] for logistic regression and 0.624 [0.587, 0.661] for the random forest. In the window of 12 months (1,631 and 460 accounts) it was 0.609 [0.580, 0.638] and 0.586 [0.557, 0.616] (Table IV). These values are similar to, or slightly higher than, those of the whole-history analysis (0.596 and 0.611), so a small difference is already present in the months before users become inactive. Four measures are lower in the attrition group in both windows after Holm correction: problems solved per contest (r = $-$0.29 and $-$0.22), struggle time (r = $-$0.17 and $-$0.12), overall acceptance rate (r = $-$0.14 and $-$0.09) and math-tag acceptance rate (r = $-$0.12 and $-$0.08). Because only 33\% (6 months) and 51\% (12 months) of attrition accounts had five rated contests in the window, this analysis describes users who were still competing regularly in the months before they stopped.

\section{Discussion}

\subsection{What the Results Show}

Users who became inactive differ from active users of similar peak rating in a few measurable ways: they accepted slightly smaller shares of their attempts and solved fewer problems per rated contest. These differences are small, persist across sampling, cleaning rules and definitions of inactivity, and shrink when tenure is controlled. When the measures are restricted to the months before inactivity, the groups still differ about as much, so part of the difference precedes inactivity; problems solved per contest remains the clearest signal. The difference is small: it does not show why users stop, and it cannot identify individual users at risk.

\subsection{Limitations}

The label is inactivity at one date. Some accounts labeled attrition may have resumed after $T_{\mathrm{obs}}$, and longer inactivity thresholds reduce but do not remove this error. A re-query of the public Codeforces API on 17 November 2026, about five months after $T_{\mathrm{obs}}$, found that 296 of the 908 accounts labeled attrition (32.6\%) had submitted again; none had been deleted. The label therefore marks a pause of at least three months at $T_{\mathrm{obs}}$, not a permanent departure.

The retrieved histories of 888 accounts were incomplete, all of them active. They were excluded, but the cause is not yet established, and the remaining imbalance in submissions per contest suggests that some incomplete active histories may remain. The rule for excluding them was chosen after inspecting the data; Table IV shows how stricter rules change the results.

The data come from one platform and from a worldwide population of users, not only students, so the results do not describe any particular group of learners. The observation time $T_{\mathrm{obs}}$ was not recorded by the collection code and is reconstructed from the data (Section III). The analysis of activity before inactivity uses histories retrieved again in November 2026 and includes only accounts with enough rated contests in the window. The two groups differ in tenure, and matching on tenure reduces the signal. Finally, the analysis reports several variants of the same models; although the two models were fixed in advance, the variants are not independent tests and should be read as a sensitivity analysis.

\section{Conclusion}

In a rating-restricted cohort of 4,329 Codeforces accounts, users who became inactive accepted slightly smaller shares of their attempts and solved fewer problems per contest than active users of similar skill. Models built on eight window-invariant measures separate the groups only weakly (ROC-AUC 0.58–0.66 across analyses) and give poorly informative probabilities. A difference of similar size is already present in the months before inactivity. Platform logs alone therefore offer a modest description of disengagement, not a means of predicting it for individuals. The archived data and code allow every result to be checked and extended.

\appendices

\section{Feature Definitions}

\noindent Table V lists every measure, its definition as implemented in the collection code, and its use in this paper.

\begin{table*}[!t]\centering\caption{Behavioral measures as implemented}\label{tab:5}

\footnotesize\setlength{\tabcolsep}{4pt}

\begin{tabular}{@{}L{0.245\dimexpr \textwidth-6\tabcolsep\relax}L{0.439\dimexpr \textwidth-6\tabcolsep\relax}L{0.194\dimexpr \textwidth-6\tabcolsep\relax}L{0.122\dimexpr \textwidth-6\tabcolsep\relax}@{}}

\toprule Measure & Definition & Source & Use \\ \midrule

Problems solved per contest & Distinct problems solved / rated contests entered, (1) & API submissions, contests & model \\

Math-tag acceptance rate & Accepted / all submissions on problems tagged math, number theory, combinatorics or probabilities (overall rate if none) & API submissions & model \\

Data-structure acceptance rate & Accepted / all submissions on problems tagged data structures, trees, graphs or dsu (overall rate if none) & API submissions & model \\

Overall acceptance rate & Accepted submissions / all submissions & API submissions & model \\

Struggle time & Mean minutes from first attempt to first acceptance, over solved problems & API submissions & model \\

Upsolves per contest & Accepted practice-mode submissions on problems of contests the user was rated in / rated contests & API submissions & model \\

Post-drop recovery & Mean rating change in the contest after a rating drop, in rating points (0 if no drop) & API rating history & model \\

Editorial dependency index & Upsolves per contest $\times$ struggle time, (2) & constructed & model \\

Rated contests per month & Rated contests / months from first to last submission (at least 1) & API contests & descriptive \\

Problems solved per month & Distinct problems solved / months from first to last submission (at least 1) & API submissions & descriptive \\

Account age & Months from registration to $T_{\mathrm{obs}}$ & API profile & descriptive \\

Rated contests (total) & Number of rated contests entered & API rating history & descriptive \\

Submissions per contest & Retrieved submissions / rated contests & API submissions & data quality \\

\bottomrule\end{tabular}

\par\vspace{3pt}\begin{minipage}{\textwidth}\raggedright\scriptsize A month is 30 days throughout. Inactivity: ($T_{\mathrm{obs}}$ $-$ last submission) / 30 days $\geq$ 3, applied to the unrounded value.\end{minipage}

\end{table*}

\section*{Data and Code Availability}

\noindent The pseudonymized Codeforces feature file from which the analysis cohort is derived, all analysis code and a script that regenerates every number, table and figure of this paper (reproduce\_all.sh) are archived on Zenodo:

\begin{center}\url{https://doi.org/10.5281/zenodo.24054464}\end{center}

\noindent and developed in the public repository

\begin{center}\url{https://github.com/AZUAD-ISLAM-RUHAN/cp-attrition-prediction}\end{center}

\noindent The raw API responses contain Codeforces handles, which identify users; they, the histories retrieved again in November 2026 for the analysis of activity before inactivity, and the salt used to pseudonymize handles are not public, and are available on reasonable request for verification.

\section*{Acknowledgment}

An earlier version of this work was presented orally at EFAST 2026. We are grateful to the anonymous reviewers and the AJSE Editorial Desk for their detailed scrutiny.

\section*{Use of AI Tools}

ChatGPT (OpenAI) was used for language editing, to refine the research topic and the initial proposal, and to help draft parts of the original data-collection and analysis code, which the authors reviewed and ran. ChatGPT also suggested references for the Related Work section of version 1; three of them did not exist and were not checked by the authors. Every reference in this version has been checked against Crossref or DataCite. Grammarly was used for language editing. In preparing this version, Claude Opus 5.5 (Anthropic) was used to write and debug the analysis, figure and manuscript-generation code and its verification tests, to check every reference against Crossref, and to draft text; OpenAI tools were also used to organize the \LaTeX{} source and to debug code. The authors ran every script, checked the reported numbers against the outputs, and take full responsibility for the content.

\end{document}